\documentclass{kapproc} 

\usepackage{procps}

\usepackage[dvips]{graphicx}

\upperandlowercase

\kluwerbib 

\usepackage{epsfig}

\begin{document}

\articletitle[Massive galaxies at $z=2$]
{Massive galaxies at redshift 2 in \\
cosmological hydrodynamic simulations
}

\author{Kentaro Nagamine,\altaffilmark{1} Renyue Cen,\altaffilmark{2}
Lars Hernquist,\altaffilmark{3} Jeremiah P. Ostriker,\altaffilmark{2}
and Volker Springel\altaffilmark{4}}

\affil{
\altaffilmark{1}University of California, San Diego, \
\altaffilmark{2}Princeton University, \ 
\altaffilmark{3}Harvard University, \\
\altaffilmark{4}Max-Planck-Institut f\"{u}r Astrophysik}


\begin{abstract}
We study the properties of galaxies at redshift $z=2$ in a $\Lambda$ 
cold dark matter ($\Lambda$CDM) universe, using two different types 
of hydrodynamic simulation methods -- Eulerian TVD and smoothed 
particle hydrodynamics (SPH) -- and a spectrophotometric analysis 
in the $U_n, G, R$ filter set.  The simulated galaxies at $z=2$ 
satisfy the color-selection criteria proposed by Adelberger et al. 
(2004) and Steidel et al. (2004) when we assume Calzetti extinction 
with $E(B-V)=0.15$.  We find that the number density of simulated 
galaxies brighter than $R<25.5$ at $z=2$ is about 
$1\times 10^{-2}~h^3~{\rm\,Mpc}^{-3}$ for $E(B-V)=0.15$, which is 
roughly twice that of the number density found by Erb et al. (2004) 
for the ultraviolet (UV) bright sample. This suggests that roughly 
half of the massive galaxies with 
${\rm M_{\star}}>10^{10}h^{-1}{\rm M_\odot}$ at $z=2$ are UV bright 
population, and the other half is bright in the infra-red (IR) 
wavelengths. The most massive galaxies at $z=2$ have stellar masses 
$\geq 10^{11-12}{\rm M}_{\odot}$. They typically have been continuously 
forming stars with a rate exceeding $30~{\rm M}_{\odot}~{\rm yr}^{-1}$ 
over a few Gyrs from $z=10$ to $z=2$, together with significant 
contribution by starbursts reaching up to  
$1000\,{\rm M}_{\odot}\,{\rm yr}^{-1}$ which lie on top of the 
continuous component. TVD simulations indicate a more sporadic star 
formation history than the SPH simulations.  
Our results do {\it not} imply that hierarchical galaxy formation 
fails to account for the observed massive galaxies at $z\geq 1$.
The global star formation rate density in our simulations peaks 
at $z\geq 5$, a much higher redshift than predicted by the 
semianalytic models. This star formation history suggests early 
build-up of the stellar mass density, and predicts that 
70 (50, 30)\% of the total stellar mass at $z=0$ had already been 
formed by $z=1$ (2, 3). Upcoming observations by {\it Spitzer} and 
{\it Swift} might help to better constrain the star formation 
history at high redshift. 
\end{abstract}



\section{Introduction}
A number of recent observational studies have revealed a new 
population of red, massive galaxies at redshift $z\sim 2$ (e.g. 
Chen et al. 2003; Daddi et al. 2004; Franx et al. 2003; Glazebrook 
et al. 2004; McCarthy et al. 2004; van Dokkum et al. 2004), 
utilizing near-IR wavelengths which are less affected by dust 
extinction. At the same time, some studies focused on the assembly of 
stellar mass density at high redshift by comparing the observational 
data and semi-analytic models of galaxy formation (e.g. Dickinson 
et al. 2003; Fontana et al. 2003; Poli et al. 2003). These works have 
suggested that the hierarchical structure formation theory may have 
difficulty in accounting for sufficient early star formation.  
The concern grew with the mounting evidence for high redshift galaxy 
formation including the discovery of Extremely Red Objects (EROs) at 
$z\ge 1$, sub-millimeter galaxies at $z\ge 2$, Lyman break galaxies 
(LBGs) at $z\geq 3$, and Lyman-$\alpha$ emitters at $z\geq 4$.
We now face the important question as to whether this evidence for
high-redshift galaxy formation is consistent with the concordance 
$\Lambda$CDM model. See Nagamine et al. (2004a,b) for the details of 
the simulations and the present work.

\begin{figure*}
\epsfig{file=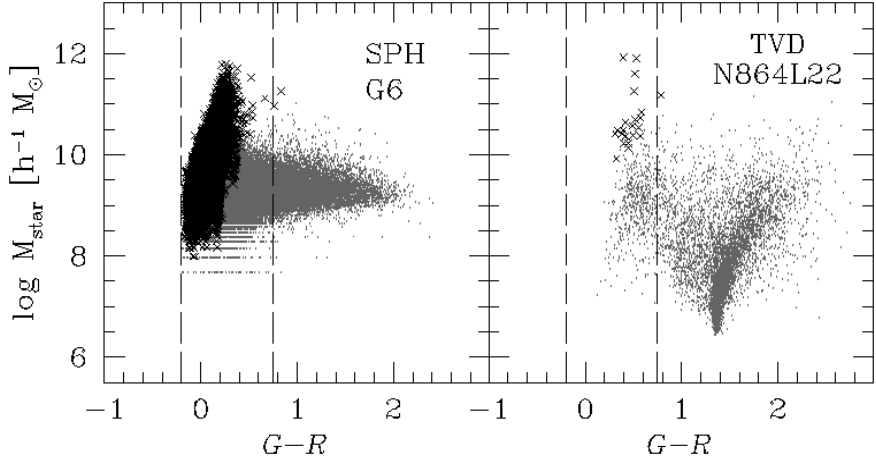,height=6cm,width=11.5cm, angle=0} 
\caption{Stellar mass vs. $G-R$ color of simulated galaxies at $z=2$.
Gray points show the total galaxy population, and the black crosses 
indicate the galaxies brighter than $R=25.5$. The vertical 
long-dashed lines roughly indicate the UV color selection range of 
Steidel et al. (2004).}
\label{f1.eps}
\end{figure*}


\section{Results}

Figure 1 shows the stellar mass vs. {\it G-R} color of the simulated 
galaxies at $z=2$. The most massive galaxies have stellar masses 
${\rm M_{\star}}>10^{10} h^{-1}M_{\odot}$, and UV colors $-1.2<G-R<0.8$, 
consistent with the UV color selection criteria of Steidel et al. 
(2004). The differences in the distribution of the points can be 
understood in terms of the different box sizes and the randomness 
of the initial condition of the simulations. On the right-hand-side 
of the panels, there are a couple of red, passive systems that have 
stellar masses ${\rm M_{\star}}>10^{10} h^{-1}{\rm M}_{\odot}$. 
The near-IR properties of these passive systems will be reported in 
future papers. Figure 2 shows the star formation histories of the 
most massive and most reddest galaxies in the simulations. For the 
most massive systems, the star formation rate has a continuous 
component of $\ge 30~{\rm M}_{\odot}~{\rm yr}^{-1}$ over a few Gyrs, 
and a starburst component exists on top of the continuous component
that reaches up to $1000\,{\rm M}_{\odot}\,{\rm yr}^{-1}$. Such extreme 
star formation histories allow these galaxies to build up stellar 
masses larger than a few $\times 10^{11}\,{\rm M}_{\odot}$ by $z=2$.
The TVD simulation indicates somewhat more sporadic star formation 
history, which is perhaps due to the differences in the details of 
the star formation recipe and the resolution. 

\begin{figure*}
\begin{center}
\resizebox{5.8cm}{!}{\includegraphics{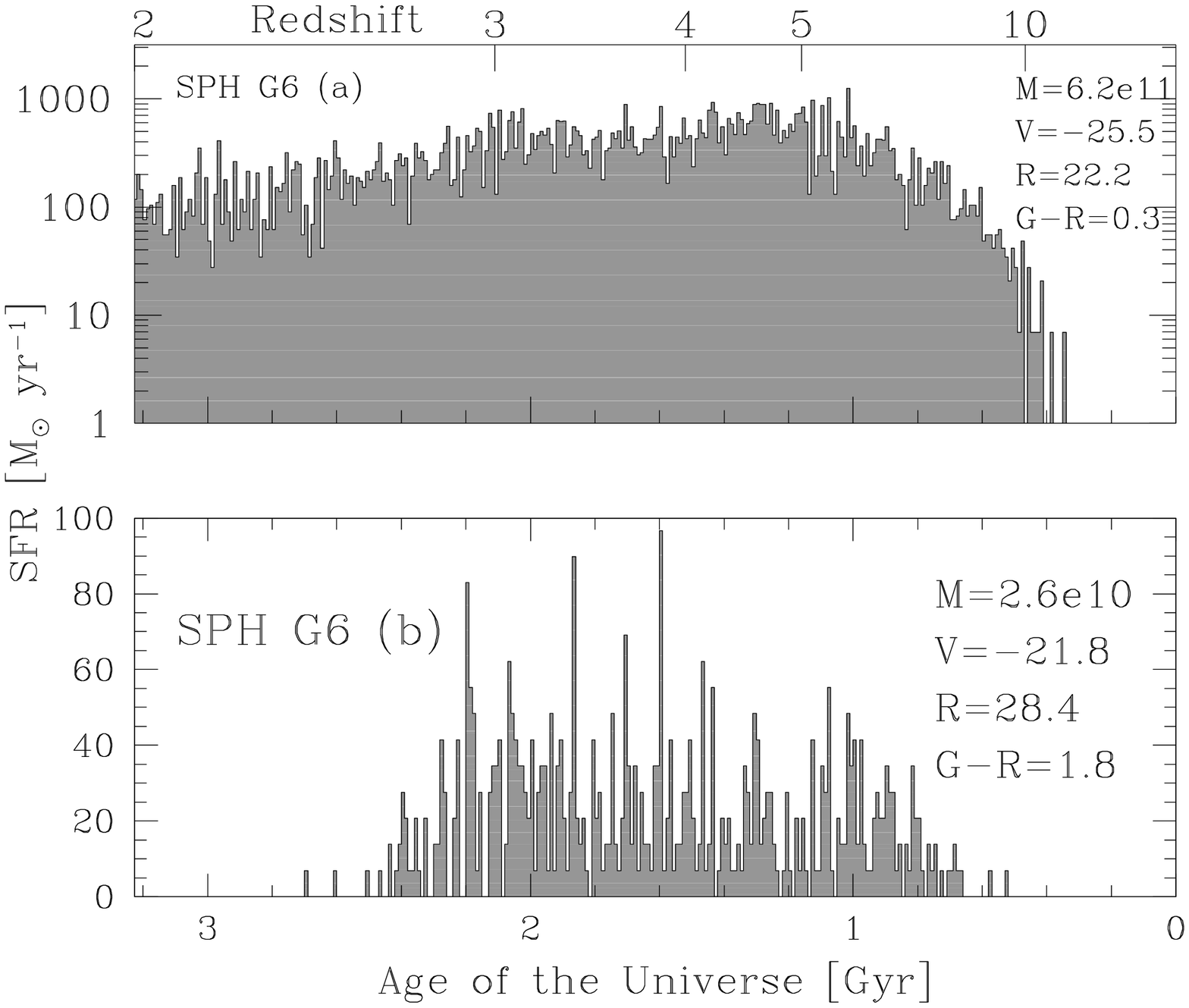}}%
\hspace{0.2cm}
\resizebox{5.8cm}{!}{\includegraphics{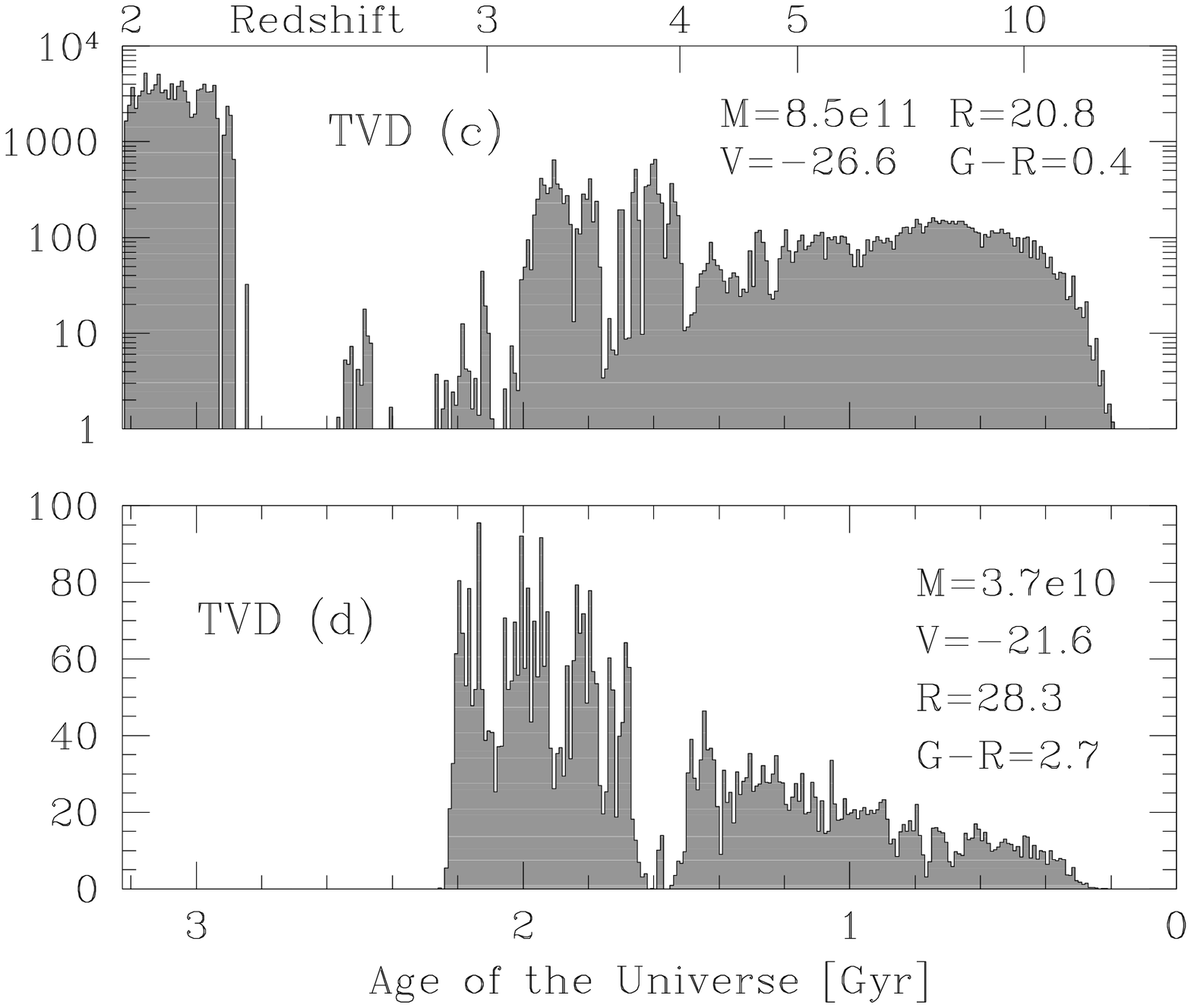}}\\%
\caption{Star formation history of the most massive galaxies 
(panels (a) \& (c)) and the reddest galaxies (panels (b) \& (d)) 
in the simulation. The left two panels are for the SPH G6 run, and 
the right ones are for the TVD run.}
\label{f2.eps}
\end{center}
\end{figure*}

Figure 3 shows the star formation history of the entire simulation 
box. In panel (a), all the models including the analytic model of 
Hernquist \& Springel (2003) show that the SFR density peaks at 
$z\ge 5$. These SFR histories lead to early build-up of the stellar 
mass density compared to both the current observational estimates and 
the results of the semianalytic models (see Nagamine et al. 2004b for 
a direct comparison), and we predict that 70 (50, 30)\% of the total 
stellar mass at $z=0$ had already been formed by $z=1$ (2, 3) based 
on these theoretical models. 

In summary, we have shown that the simulations based on the 
hierarchical CDM model can in fact account for the masses and the 
comoving number densities of the massive galaxies at $z=2$ that are 
found by the recent observations. Our simulations indicate that the 
properties (i.e. stellar mass, color, SF history, clustering) of the 
UV bright LBGs at $z\ge 2$ can be understood if they are identified with 
the galaxies that reside in the massive dark matter halos at the time.

\begin{figure*}
\begin{center}
\resizebox{5.8cm}{!}{\includegraphics{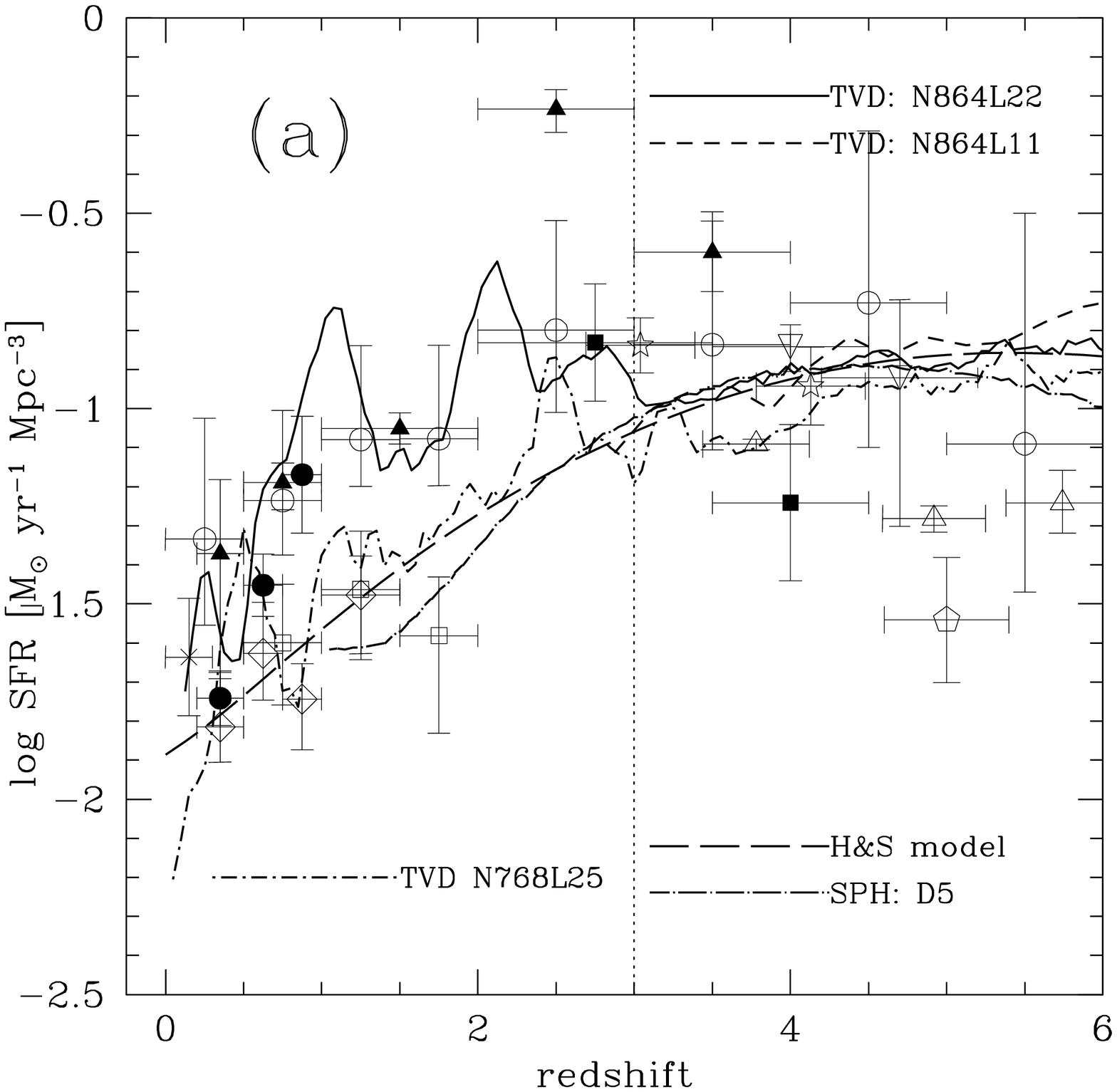}}%
\hspace{0.2cm}
\resizebox{5.8cm}{!}{\includegraphics{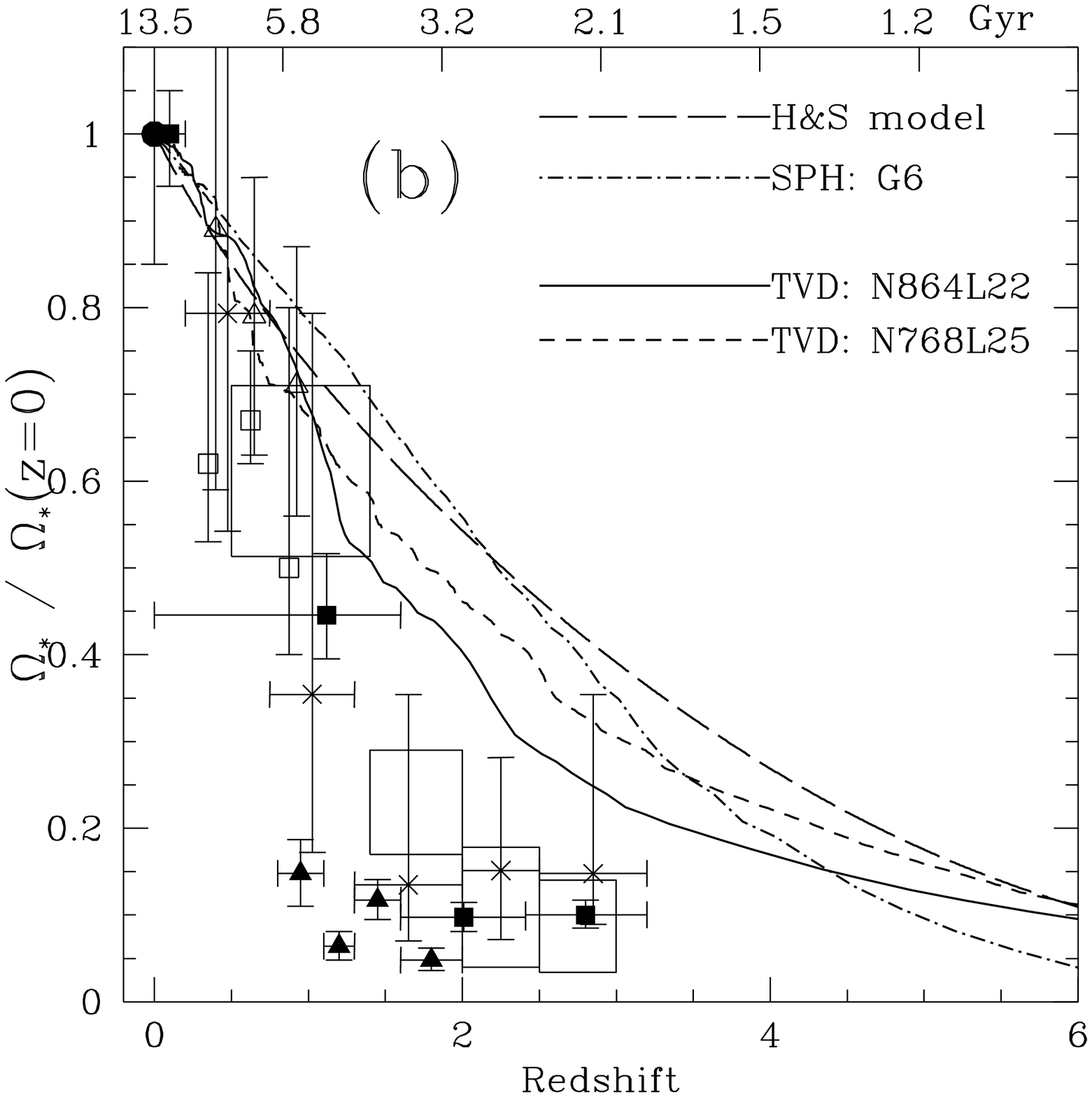}}\\%
\caption{Panel (a): Star formation rate density as a function of redshift for simulations
and the analytic model by Hernquist \& Springel (2003).
The sources of the extinction corrected data points are described in Nagamine et al. (2004a).
Panel (b): Stellar mass density as a function of redshift.
}
\label{f3.eps}
\end{center}
\end{figure*}

\begin{acknowledgments}
This work was supported by NSF grants ACI 96-19019, AST 00-71019, AST 02-06299,
and AST 03-07690, and NASA ATP grants NAG5-12140, NAG5-13292, and NAG5-13381.
\end{acknowledgments}

\end{document}